\documentstyle[11pt,ask,epsf]{article}
\epsfverbosetrue
\newcommand{\IRSaone}%
{\begin{picture}(12,4)
\put(0,0){\line(1,0){12}}
\put(0,4){\line(1,0){12}}
\put(0,0){\line(0,1){4}}
\put(4,0){\line(0,1){4}}
\put(8,0){\line(0,1){4}}
\put(12,0){\line(0,1){4}}
\end{picture}}
\newcommand{\IRSe}%
{\begin{picture}(8,8)
\put(0,0){\line(1,0){4}}
\put(0,4){\line(1,0){8}}
\put(0,8){\line(1,0){8}}
\put(0,0){\line(0,1){8}}
\put(4,0){\line(0,1){8}}
\put(8,4){\line(0,1){4}}
\end{picture}}
\newcommand{\IRSatwo}%
{\begin{picture}(4,12)
\put(0,0){\line(1,0){4}}
\put(0,4){\line(1,0){4}}
\put(0,8){\line(1,0){4}}
\put(0,12){\line(1,0){4}}
\put(0,0){\line(0,1){12}}
\put(4,0){\line(0,1){12}}
\end{picture}}
\makeatletter
\@addtoreset{equation}{section}
\makeatother

\def\veg#1{{\bf #1}}
\def\vek#1{\mbox{\protect\boldmath $#1$}}
\newcommand{\Z}{{\sf Z \!\!\! Z}}
\newcommand{\R}{\mbox{\sf I \hspace*{-0.75em} R}}

\newcommand{\cD}{{\cal D}}
\newcommand{\cG}{{\cal G}}
\newcommand{\cL}{{\cal L}}
\newcommand{\cS}{{\cal S}}
\newcommand{\cV}{{\cal V}}
\def\diag{\mathop{\rm diag}}
\def\gsim{\;\raisebox{-.4ex}{\rlap{$\sim$}} \raisebox{.4ex}{$>$}\;}
\newcommand{\half}{\mbox{\small $\frac{1}{2}$}}

\newcommand{\quart}{\mbox{\small $\frac{1}{4}$}}

\newcommand{\third}{\mbox{\small $\frac{1}{3}$}}

\newcommand{\Tr}{{\rm Tr}}

\preprint{FERMILAB-PUB-92/255-T \\ BUTP-92/40}
\title{SU($N$) Gauge Theories with $C$-Periodic Boundary Conditions:
II. Small Volume Dynamics}
\author{A.S. Kronfeld$^1$ and
U.-J. Wiese$^2$\thanks{supported by the Schweizer Nationalfond} \\[2.0em]
$^1$Theoretical Physics Group, Fermi National Accelerator
Laboratory,\thanks{Fermilab is operated by Universities Research
Association Inc.\ under contract with the U.S. Department of Energy.}
 \\ P.O. Box 500, Batavia, IL 60510, USA \\[0.5em]
$^2$Universit\"at Bern, Sidlerstrasse 5, 3012 Bern, Switzerland}
\date{October 6, 1992}
\begin{document}
\maketitle
\vfill
\begin{abstract} \normalsize
The dynamics of SU($N$) gauge theories, especially for $N=3$, in a small
$C$-periodic box are investigated.
We identify the fields that mimimize the energy---the torons---and
determine which of these ``classical'' vacua are stable quantum
mechanically.
The stable torons break cubic symmetry, which has interesting
consequences on the spectrum.
At any of the stable torons there are also quartic modes.
Since all $C$-periodic boundary conditions are gauge-equivalent, we
choose a convenient version, for which the quartic modes are constant
modes, and compute the effective Hamiltonian to one loop in perturbation
theory.
\end{abstract}
\vfill \newpage \setcounter{page}{1} \pagestyle{plain}
\section{Introduction}
For non-abelian gauge fields $C$-periodic boundary conditions were
introduced in ref.~\cite{Kro91}.
With $C$-periodic boundary conditions fields at $\vek{x}+L\vek{e}_i$ are
the charge conjugate of fields at $\vek{x}$.
(Here $\vek{e}_i$ is a unit vector in the $i$ direction, and $L$ denotes
the length of the torus.)
Gauge invariant operators with $C=+1$ are periodic and
those with $C=-1$ are anti-periodic.
An SU($N$) gauge field, however, is $C$-periodic up to a gauge
transformation:
\begin{equation}\label{cpbc}
\vek{A}(\vek{x}+L\vek{e}_i) = \Omega_i(\vek{x})
[\vek{\nabla} + \vek{A}^{*}(\vek{x}) ] \Omega_i^{-1}(\vek{x}),
\end{equation}
where $*$ denotes complex conjugation.
(For details about our conventions, see \cite{Kro91}.)
Under local gauge transformations
\begin{equation}\label{gA}
^g\vek{A}(\vek{x}) =
g(\vek{x}) [\vek{\nabla} + \vek{A}(\vek{x}) ] g^{-1}(\vek{x}),
\end{equation}
and
\begin{equation}\label{gOmega}
^g\Omega_i(\vek{x}) =
g(\vek{x}+L\vek{e}_i) \Omega_i(\vek{x}) g^{\rm T}(\vek{x}),
\end{equation}
where $g(\vek{x})$ is an SU($N$)-valued function with arbitrary
boundary condition.
The superscript ${\rm T}$ denotes transpose, and for a unitary matrix
$g^{\rm T}=(g^{-1})^{*}$.

$C$-periodic boundary conditions are interesting because they share
several topological properties of the infinite volume
\cite{Kro91,Pol91}.
For example, in Abelian gauge theories $C$-periodic boundary conditions
permit a single charged particle in a finite volume, whereas periodic
systems are always neutral \cite{Pol91}, owing to Gauss' law.
In pure SU($N$) gauge theories with periodic boundary conditions
the Hamiltonian possesses $\Z_N$ symmetries.
With $C$-periodic boundary conditions there are no such symmetries for
odd $N$, and they are merely $\Z_2$ for even $N$ \cite{Kro91}.
Furthermore, for odd $N$ the topological (instanton) charge is an
integer.

At non-zero temperature, spatial $C$-periodic boundary conditions
explicitly break the (``temporal'') $\Z_3$ symmetry of the SU(3)
partition function.
They act as a soft source, selecting a preferred value of the Polyakov
loop \cite{Wie91}.
This contrasts the behavior in a periodic box, where the average
Polyakov loop vanishes because of tunneling between the three $\Z_3$
phases, even above the deconfinement temperature.
Physically speaking, a single $\Z_3$ charge---the Polyakov loop
represents a static quark---can exist in a $C$-periodic volume, while in
a periodic volume the $\Z_3$ Gauss law forbids it.
When dynamical quarks are included $C$-periodic boundary conditions
break chiral symmetry in a similar fashion \cite{Wie91}, which has
interesting consequences for the low energy chiral Lagrangian describing
the physics of the Goldstone pions.
More generally, $C$-periodic boxes offer the possibility of approaching
the infinite volume limit with different qualitative features than in
traditional periodic boxes.
In particular, a comparison of the two can be used to estimate the size
of finite size effects, and to show that the physics ultimately
becomes independent of the infrared cutoff.

This paper concentrates on $N=3$ and studies the dynamics in a small
$C$-periodic box using perturbation theory.
For the periodic box this approach was pioneered by L\"uscher
\cite{Lue83}.
To all orders of perturbation theory, he found that the spectrum
possesses an $N^d$-fold degeneracy among all sectors of 't~Hooft
electric flux \cite{tHo79}.
The degeneracy is lifted non-perturbatively by tunneling between
different perturbatively degenerate states \cite{Kol86}.
For SU(3) gauge theory with $C$-periodic boundary conditions there are
no $\Z_3$ sectors at all.
Therefore, one might hope that the spectrum in a $C$-periodic box
does not get rearranged by tunneling, and that the approach to the
infinite volume limit might be smoother than in the periodic case.
Unfortunately, it turns out that states with different quantum numbers
of the cubic rotation group are degenerate to all orders of perturbation
theory.
Once again, non-perturbative tunneling removes the degeneracy.

We follow L\"uscher's line of attack \cite{Lue83}.
First, one identifies the configurations that minimize the potential
energy---the so-called torons.
Second, one integrates out all modes other than the torons, yielding an
effective potential $\cV_{\rm eff}$, the minima of which are the
starting point of perturbation theory.
As with periodic boundary conditions, it turns out that there
are other modes (other than the torons) that are quartic
at the minima of $\cV_{\rm eff}$.
They cannot be integrated out perturbatively, and the small-volume
dynamics is described by an effective Hamiltonian $H_{\rm eff}$
for the quartic modes and the torons.
Note that the two effective descriptions apply in different regions of
configuration space, as illustrated in fig.~\ref{expansions}.
\begin{figure}[b]
\epsfxsize=\textwidth
\epsfbox{cpbcII1.eps}
\caption[expansions]{Regions of validity of $\cV_{\rm eff}$ vs.\
$H_{\rm eff}$.
The modes corresponding to the direction labeled ``quadratic'' are
integrated out in both cases.
For the expansions to be valid, their fluctuations must be
${\rm O}(g_0)$.
$\cV_{\rm eff}$ is valid in the gray box, when the ``quartic''
fluctuations are ${\rm O}(g_0)$ but the torons are O(1).
$H_{\rm eff}$ is valid in the clear box, when the quartic and toron
fluctuations are ${\rm O}(g_0^{2/3})$.}\label{expansions}
\end{figure}

The paper is organized as follows:
Sect.~\ref{valley} explores the structure of the $C$-periodic toron
valley, which consists of real, constant, abelian fields.
The appendices contain the detailed proof that the torons are
gauge-equivalent to real, constant, abelian fields.
Sect.~\ref{Veff} presents the calculation of the toron effective
potential $\cV_{\rm eff}$.
It has four absolute minima and one local minimum.
Any of the four absolute minima is an appropriate starting point for
perturbation theory, but they all break cubic invariance.
The consequences of this symmetry breaking on the spectrum in small
volumes is sketched in sect.~\ref{degen}.
Sect.~\ref{Heff} explains how to evaluate $H_{\rm eff}$ by evaluating
the (imaginary) time evolution amplitude and presents the results of
the one-loop approximation to $H_{\rm eff}$.
Finally, sect.~\ref{conclude} offers some concluding remarks.

\section{Vacuum valley for $C$-periodic boundary
conditions}\label{valley}
In SU(3) all choices of the transition function $\Omega_i$ in
eq.~(\ref{cpbc}) are physically equivalent \cite{Kro91}, so, except in
sect.~\ref{Heff}, we shall work with $\Omega_i=1$.
Then eq.~(\ref{cpbc}) becomes
\begin{equation}\label{component-bc}
\begin{array}{r@{\,=\,}ll}
\vek{A}^A(\vek{x}+L\vek{e}_i) &   \vek{A}^A(\vek{x}), &
        A\in\{2,5,7\}, \\[0.7em]
\vek{A}^a(\vek{x}+L\vek{e}_i) & - \vek{A}^a(\vek{x}), &
        a\in\{1,3,4,6,8\},
\end{array}
\end{equation}
in Lie-algebra components.
More generally, the periodic components take values in an so($N$)
subalgebra of su($N$), generated by the $N$ real generators.
The Hamiltonian can be written as $H=\,^+\!H +\,^-\!H$, where
\begin{equation}\label{ham-split}
\begin{array}{l}
^+\!H = {\displaystyle \int} d^dx \left[
\half g_0^2(E_i^A)^2 + \quart g_0^{-2}(F_{ij}^A)^2 \right], \\[0.7em]
^-\!H = {\displaystyle \int} d^dx \left[
\half g_0^2(E_i^a)^2 + \quart g_0^{-2}(F_{ij}^a)^2 \right].
\end{array}
\end{equation}
(We work in $d$ space dimensions, but we are most interested in $d=3$.)
The appendices show in detail that the minima of the potential energy
are gauge-equivalent to real, constant, abelian fields.
Roughly speaking, the anti-periodic modes must vanish because otherwise
$\nabla_iA^a_j\neq0$ and, hence, $F^a_{ij}\neq0$.
This leaves the real so($N$) subalgebra with Hamiltonian $^+\!H$,
whose potential $\int d^dx(F_{ij}^A)^2/(4g_0^2)$ is minimized by
constant, abelian fields.

In SU(2) and SU(3) the manifold of distinct torons resembles the
orbifold that appears for SU(2) with periodic boundary conditions.
The real, constant, abelian field is
\begin{equation}\label{toron}
^g\!\vek{A}=\frac{\vek{C}}{L}T^2,
\end{equation}
where $T^a=i\sigma^a/2$ in SU(2) and $T^a=i\lambda^a/2$ in SU(3).
At the outset, it appears that $\vek{C}\in\R^d$, but there are discrete
gauge symmetries identifying certain $\vek{C}$ \cite{Kol86}.
The transformation
\begin{equation}
\exp(-4\pi T^2x_i/L):\;C_i\mapsto C_i + 4\pi
\end{equation}
reduces the manifold to a torus.
Writing $\vek{l}=\sum_{i=1}^d\vek{e}_i$, the transformation
\begin{equation}\label{orbifold-1}
\exp(-3\pi T^2\vek{x}\cdot\vek{l}/L) \exp(\pi T^3)
\exp(-\pi  T^2\vek{x}\cdot\vek{l}/L):\;		
\vek{C}\mapsto -\vek{C} + 2\pi\vek{l} 
\end{equation}
reduces the manifold to an orbifold. 
For SU(2) the orbifold is $[0, 4\pi)^d/\Z_2$, as with periodic
boundary conditions. 
This is no great surprise, since for SU(2) $C$-periodic boundary
conditions are gauge equivalent to periodic ones \cite{Kro91}.

For SU(3) there is yet another discrete symmetry:
\begin{equation}
\exp(4\pi T^5\vek{x}\cdot\vek{l}/L):\; \vek{C}\mapsto -\vek{C}
\end{equation} 
reducing the orbifold to $[0, 4\pi)^d/(\Z_2\times\Z_2)$. 
The singular points of this orbifold are (for $d=3$) in the set
$\Pi=\{(\pi,\pi,\pi),\,(\pi,\pi,-\pi),\,
     (\pi,-\pi,\pi),\,(\pi,-\pi,-\pi)\}$.
Based on previous experience, we expect these points to play an
important role.

\section{Effective potential for $\vek{C}$}\label{Veff}

To work out the toron effective potential, we start by parametrizing the
SU(3) gauge field as follows:
\begin{equation}
\begin{array}{c}
L\vek{A}(\vek{x})= g^{-1}(\vek{x})\left( L\vek{\nabla} + \vek{C} T^2
+ g_0 {\displaystyle \sum_{\vek{\nu}}} \vek{q}^A(\vek{\nu}) T^A
                           \exp[i2\pi\vek{\nu}\cdot\vek{x}/L]
\right. \hspace*{1.25cm} \\ \hspace*{1.25cm} \left.\rule{0.em}{1.0em}
+ g_0 {\displaystyle \sum_{\vek{\nu}}} \vek{q}^a(\vek{\nu}) T^a
                           \exp[i\pi(2\vek{\nu}+\vek{l})\cdot\vek{x}/L]
\right) g(\vek{x})
\end{array}
\end{equation}
where $\vek{q}^{2}(\veg{0})=0$,
$\vek{q}^A(-\vek{\nu})=\vek{q}^A(\vek{\nu})^*$,
$\vek{q}^a(-\vek{\nu}-\vek{l})=\vek{q}^a(\vek{\nu})^*$.
Upper-case color indices run over $A\in\{2,5,7\}$, whereas
lower-case color indices run over $a\in\{1,3,4,6,8\}$.

Let us define background-field covariant derivatives
\begin{equation}
^+\vek{\cD}^{AB} = i2\pi\vek{\nu}\delta^{AB}           + \vek{C} f^{A2B}
\end{equation}
acting on $\vek{q}^A(\vek{\nu})$ and
\begin{equation}
^-\vek{\cD}^{ab} = i\pi(2\vek{\nu}+\vek{l})\delta^{ab} + \vek{C} f^{a2b}
\end{equation}
acting on $\vek{q}^a(\vek{\nu})$.
To ${\rm O}(g_0^2)$ the kinetic energy of $\vek{C}$ and interactions
can be neglected and the Hamiltonian is
\begin{equation}
^+\!H=\frac{1}{2L}
\sum_{\vek{\nu}} \left[ \vek{p}^A(-\vek{\nu})\cdot \vek{p}^A(\vek{\nu})
+ q_j^A(-\vek{\nu})\,^+\Omega_{jk}^{AB} q_k^B(\vek{\nu}) \right]
\end{equation}
with $\vek{p}^A(\vek{\nu})$ the canonical momentum conjugate to
$\vek{q}^A(\vek{\nu})$ and
\begin{equation}
^+\Omega_{jk}^{AB} =
(-\,^+\cD^2\delta_{jk}+\,^+\cD_j\,^+\cD_k)^{AB} .
\end{equation}
The expressions for $^-\!H$ and $^-\Omega^{ab}_{jk}$ are obtained by
replacing $^+\vek{\cD}^{AB}$ by $^-\vek{\cD}^{ab}$,
$\vek{q}^A$ by $\vek{q}^a$, etc.
To ${\rm O}(g_0^2)$ the Hamiltonians $^\pm\!H$ describe the
$\vek{\nu}\neq\veg{0}$ modes as a collection of harmonic oscillators
whose squared frequencies are the eigenvalues of $^\pm\Omega$.
Adding up the zero-point energy of these oscillators gives the effective
potential for $\vek{C}$:
\begin{equation}
L\cV_{\rm eff}(\vek{C}) =
\half\Tr[\,^+\Omega^{1/2}(\vek{C})] +
\half\Tr[\,^-\Omega^{1/2}(\vek{C})].
\end{equation}
It is natural to work in the background-field Coulomb gauge
\begin{equation}
^+\vek{\cD}^{AB}\cdot\vek{q}^B(\vek{\nu})=0,\;\;\;
^-\vek{\cD}^{ab}\cdot\vek{q}^b(\vek{\nu})=0.
\end{equation}
Since $\vek{C}$ parameterizes an abelian field, the background-field
covariant derivatives commute, so in Coulomb gauge
$^\pm\Omega=-\,^\pm\cD^2$.
The eigenvalues are
\begin{equation}\label{eigen+}
^+\Omega(\vek{C}): \begin{array}{ll}
(2\pi\vek{\nu})^2,\;\; \vek{\nu}\neq0, & A=2~{\rm sector}\\
(2\pi\vek{\nu}\pm\half\vek{C})^2, & A\in\{5,7\}~{\rm sector}\end{array}
\end{equation}
and
\begin{equation}\label{eigen-}
^-\Omega(\vek{C}): \begin{array}{ll}
(2\pi\vek{\nu}+\pi\vek{l}\pm\vek{C})^2, & a\in\{1,3\}~{\rm sector} \\
(2\pi\vek{\nu}+\pi\vek{l}\pm\half\vek{C})^2, & a\in\{4,6\}~{\rm sector} \\
(2\pi\vek{\nu}+\pi\vek{l})^2, & a=8~{\rm sector} \end{array}
\end{equation}
all of which have multiplicity $d-1$.
Performing the traces yields
\begin{equation}\label{V-traces}
\cV_{\rm eff}(\vek{C}) = \hat{\cV}(\half\vek{C}) +
\hat{\cV}(\vek{C}-\pi\vek{l}) +
\hat{\cV}(\half\vek{C}-\pi\vek{l})
\end{equation}
where \cite{Lue83,Kol86}
\begin{equation}
\hat{\cV}(\vek{C})=-\frac{d-1}{L\pi^2}\sum_{\vek{\nu}\neq\veg{0}}
\frac{1}{\nu^4}\cos(\vek{C}\cdot\vek{\nu}).
\end{equation}
Near its minimum at $\vek{C}=\veg{0}$,
$\hat{\cV}(\vek{C})\approx\hat{\cV}(\veg{0})+2|\vek{C}|/L$, cf.\
ref.~\cite{Kol86}.
Hence, the effective potential $\cV_{\rm eff}(\vek{C})$ has conical
minima at $\vek{C}=\veg{0}$, from the first term in
eq.~(\ref{V-traces}), and at the $2^{d-1}$ points in $\Pi$, from the
second term.
(The third term implies a minimum at $2\pi\vek{l}$ which is equivalent
to $\veg{0}$ by eq.~(\ref{orbifold-1}).)
To decide which point is the absolute minimum one must
perform the sums explicitly.
For $d=3$
\begin{equation}
\begin{array}{r@{\,=\,}l}
L\cV_{\rm eff}(\veg{0})       & -1.78447\cdots ,\\[0.5em]
L\cV_{\rm eff}(\vek{C}\in\Pi) & -3.25229\cdots .
\end{array}
\end{equation}
Thus, there are four degenerate, absolute minima at
$\vek{C}\in\Pi$, as well as a local minimum at $\vek{C}=\veg{0}$.

Any of the four minima in $\Pi$ is an appropriate starting point for
perturbation theory.
It may seem peculiar that $\vek{C}=\veg{0}$ is {\em not\/} appropriate,
but it has precedent \cite{vBa88}.
In sect.~\ref{Heff} we shall set up perturbation theory, expanding
around $\vek{C}=(\pi,\pi,\pi)$.
The next section discusses, in general terms, how the cubic symmetry
breaking influences the spectrum.

\section{Degeneracies to all orders in perturbation theory}\label{degen}
Let us review the structure of the effective potential for SU($N$) gauge
theories with periodic boundary conditions.
The symmetry group of the Hamiltonian is $G=\Z_N^d\cdot O(d,\Z)$, the
semi-direct product of the central conjugations and the rotation
group of the $d$-cube.
The effective potential for the torons has $N^d$ minima, each with
stability subgroup $\cG=O(d,\Z)$.
If one calculates the spectrum perturbatively, by expanding about a
given minimum, one finds that the spectrum is labeled by the irreducible
representations of $\cG$.
But there is an $N^d$-fold degeneracy, coming from states localized
at the other minima.
That is why the 't~Hooft electric flux sectors are degenerate with the
glueballs, to all orders in perturbation theory.
Of course, non-perturbative tunneling between the $N^d$ minima removes
the degeneracies.

There are analogous phenomena for $N=3$ with $C$-periodic boundary
conditions.
Now the symmetry group of the Hamiltonian is $G=O(d,\Z)$, because there
are no central conjugations.
The effective potential for the torons has $2^{d-1}$ minima; below we
identify the stability group $\cG$.
Again, the irreducible representations of $\cG$ label the perturbative
spectrum, but each multiplet is $2^{d-1}$-fold degenerate.
Combining $2^{d-1}$ degenerate $\cG$-multiplets into a (reducible)
representation of $G$ then reveals that the Hamiltonian has
degeneracies, to all orders in perturbation theory.
And again, tunneling removes them.

To work out the details one needs to describe the states enough to
understand their symmetry properties.
The quantum mechanical system at hand contains the degrees of freedom
$\vek{C}$ governed by the potential $\cV_{\rm eff}(\vek{C})$.
A ``perturbative eigenstate'' centered at minimum $\vek{C}^\pi\in\Pi$
obeys
\begin{equation}
\langle \vek{C} \rangle = \vek{C}^\pi T^2/L.
\end{equation}
In this sense $\vek{C}^\pi$ characterizes the state.
Consider the Wilson loops $\exp(\oint\vek{C}^\pi\cdot d\vek{x})$,
and define
\begin{equation}
\cL_{ij}^{\pm} = \half\left[
\Tr \left\{\exp[T^2(C_i^\pi\pm C_j^\pi)]\right\} - 1\right]=
\cos\left\{ \half  (C_i^\pi\pm C_j^\pi) \right\}.
\end{equation}
Up to trivialities $\cL_{ij}^\pm$ describes a loop that wraps around the
$C$-periodic box first in the $i$ direction and then in the $\pm j$
direction, evaluated at $\vek{C}=\vek{C}^\pi$.

Table~\ref{L} contains the values of the $\cL_{ij}^{\pm}$
at all points in $\Pi$ for $d=3$.
\begin{table}
\caption[L]{Values of $\cL_{ij}^{\pm}$ for all $\vek{C}\in\Pi$
for $d=3$.  Note that, at these points,
$\cL_{ij}^{-}=-\cL_{ij}^{+}$ always.}\label{L}
\vspace{1.0em}
\begin{center} \begin{tabular}{ccccc}
\hline \hline
 & $(\pi,\pi,\pi)$  & $(\pi,\pi,-\pi)$
 & $(\pi,-\pi,\pi)$ & $(\pi,-\pi,-\pi)$ \\
\hline
$\cL_{12}^{+}$ & $-$ & $-$ & $+$ & $+$ \\
$\cL_{13}^{+}$ & $-$ & $+$ & $-$ & $+$ \\
$\cL_{23}^{+}$ & $-$ & $+$ & $+$ & $-$ \\
$\cL_{12}^{-}$ & $+$ & $+$ & $-$ & $-$ \\
$\cL_{13}^{-}$ & $+$ & $-$ & $+$ & $-$ \\
$\cL_{23}^{-}$ & $+$ & $-$ & $-$ & $+$ \\
\hline \hline
\end{tabular} \end{center}
\end{table}
At $(\pi,\pi,\pi)$ one sees that the stability subgroup is $\cG=S_3$, the
permutation group of the $d=3$ axes,\footnote{In $d$ dimensions,
$\cG=S_d$.} because any rotation that is {\em not\/} a permutation
transforms a negative $\cL$ into a positive one.
At the other points $\cG$ is isomorphic to $S_3$.
There are three irreducible representations of $S_3$, whose
Young tableaux are $\IRSaone$, $\IRSe$, and $\IRSatwo$.
If $\rho$ is one of these representations, let $R_\rho$ be the reducible
representation of $O(3,\Z)$ generated by combining $2^{d-1}=4$
representations $\rho$, one from each minimum, each with the same
(perturbative) energy.
Using character tables it is easy to work out the irreducible content of
$R_\rho$.
One finds
\begin{equation}\label{degenerate}
\begin{array}{r@{\,=\,}l}
R_{\IRSaone}    & A_1 \oplus T_2,            \\[0.5em]
R_{\IRSe}       &  E  \oplus T_1 \oplus T_2, \\[0.5em]
R_{\IRSatwo}    & A_2 \oplus T_1.
\end{array}
\end{equation}
Eqs.~(\ref{degenerate}) have the following consequences on the spectrum.
In small volumes, where tunneling effects are suppressed,
each $A_1$ state is degenerate with a $T_2$ state,
each $A_2$ state is degenerate with a $T_1$ state, and
each $E$ state is degenerate with a $T_1$ {\em and\/} a $T_2$ state.
As with periodic boundary conditions, there is presumably a tunneling
transition for some intermediate value of the coupling (i.e.\ volume)
splitting these degeneracies.

At larger volumes $L\gsim 1$~fm rotational invariance ought to be
restored.
That means that an $E$ and a $T_2$ ought to combine into a spin 2
multiplet, an $A_2$, a $T_1$, and a $T_2$ ought to combine into a spin
3 multiplet, etc.
In addition, there may be other features of the spectrum arising from
tunneling to the local minimum at $\vek{C}=\veg{0}$.

\section{The Effective Hamiltonian}\label{Heff}
At $\vek{C}=(\pi,\pi,\pi)$ the quadratic term in $^-\!H$ vanishes
for two modes,
\begin{equation}\label{quartic-modes}
\begin{array}{l}
(\vek{q}^1(\veg{0})-i\vek{q}^3(\veg{0}))/\sqrt{2},   \\[0.5em]
(\vek{q}^1(\veg{0})-i\vek{q}^3(\veg{0}))^*/\sqrt{2}=
(\vek{q}^1(-\vek{l})+i\vek{q}^3(-\vek{l}))/\sqrt{2},
\end{array}
\end{equation}
cf.\ eq.~(\ref{ham-split}).
The fluctuations of these modes and $\vek{C}$ are still
damped by quartic terms, so they are called quartic modes.
Since perturbation theory develops a series of corrections to quadratic
interactions, one must obtain an effective Hamiltonian for {\em all\/}
quartic modes and solve it non-perturbatively, as in
refs.~\cite{Lue84,Kol86}.

Expanding around $\vek{C}=(\pi,\pi,\pi)$ is a nuisance, especially since
the quartic modes in eq.~(\ref{quartic-modes}) prefer a complex basis
for the Lie algebra.
The gauge transformation
\begin{equation}\label{C-shift}
h(\vek{x})=\exp(\pi T^2\vek{x}\cdot\vek{l}/L)
\end{equation}
eliminates these problems, but induces a non-unit twist:
$^h\Omega_i=\Omega$, where
\begin{equation}
\Omega= \exp(\pi T^2) = \left(\begin{array}{rrr}
 0 & 1 & 0 \\
-1 & 0 & 0 \\
 0 & 0 & 1 \end{array}\right),
\end{equation}
as required by eq.~(\ref{gOmega}).
The transformed gauge potential obeys a new boundary condition,
\begin{equation}\label{cpbcO}
\vek{A}(\vek{x}+L\vek{e}_i) = \Omega \vek{A}^{*}(\vek{x}) \Omega^{-1},
\end{equation}
or, in explicit Lie-algebra components,
\begin{equation}\label{cpbcL}
\begin{array}{c}
\vek{A}^A(\vek{x}+L\vek{e}_i) = +\vek{A}^A(\vek{x}),\;\;A\in\{1,2,3\},
\\[0.7em]
\vek{A}^8(\vek{x}+L\vek{e}_i) = -\vek{A}^8(\vek{x}) ,\\[0.7em]
(\vek{A}^5 \pm i\vek{A}^7)(\vek{x}+L\vek{e}_i) = \pm i
(\vek{A}^5 \pm i\vek{A}^7)(\vek{x}) ,\\[0.7em]
(\vek{A}^4 \mp i\vek{A}^6)(\vek{x}+L\vek{e}_i) = \pm i
(\vek{A}^4 \mp i\vek{A}^6)(\vek{x}).
\end{array}
\end{equation}
With eq.~(\ref{cpbcO}) the torons are still real, constant, abelian
fields, parameterized as in eq.~(\ref{toron}).
Splitting the Hamiltonian into pieces according to the boundary
conditions of eq.~(\ref{cpbcL}) and repeating the analysis of
sect.~\ref{Veff} shows that the absolute minima of $\cV_{\rm eff}$ are
now at $\vek{C}=\veg{0}$ and three other points $2\pi\vek{e}_i$.
Furthermore, at $\vek{C}=\veg{0}$ the additional quartic modes are the
constant modes of $\vek{A}^1$ and $\vek{A}^3$.
Hence, the split between quartic and quadratic modes is simply
\begin{equation}
L\vek{A}^A(\vek{x})=\vek{c}^A + g_0\vek{q}^A(\vek{x}),
\;\;\;A\in\{1,2,3\},
\end{equation}
with $\int d^dx\,\vek{q}^A(\vek{x})=0$.

The components $\vek{A}^a$, $a\in\{4,5,6,7,8\}$, are always
quadratic.
$\vek{A}^8$ is anti-periodic and has half-integer momenta
$\vek{p}=2\pi(\vek{\nu}+\half\vek{l})/L$.
However, it does not couple to the quartic modes $\vek{c}^A$, so it plays no
role in the calculations below.
The $\pm i$-periodic modes have quarter-integer momenta
$\vek{p}=2\pi(\vek{\nu}\pm\quart\vek{l})/L$.
These are not artifacts of the boundary condition eq.~(\ref{cpbcO});
instead, they would have appeared with the standard $C$-periodic
boundary condition, when expanding around $\vek{C}=(\pi,\pi,\pi)$,
as a glance at the eigenvalues in eqs.~(\ref{eigen+}) and (\ref{eigen-})
verifies.

The effective Hamiltonian can be worked out by computing the amplitude
to propagate from one constant field $\vek{c}^A$ to another.
It is defined by
\begin{equation}
\left\langle\vek{c}^A(T)|e^{-H_{\rm eff}T}|\vek{c}^A(0)\right\rangle=
\left\langle\vek{c}^A(T)|e^{-HT}|\vek{c}^A(0)\right\rangle,
\end{equation}
where the right-hand side is computed in the full theory, and the
left-hand side is an expression involving $\vek{c}^A$ and its conjugate
momentum $\vek{e}^A=-i\partial/\partial\vek{c}^A$.
In the path integral formalism
\begin{equation}
\left\langle\vek{c}^A(T)|e^{-HT}|\vek{c}^A(0)\right\rangle=
\int\cD A^a_\mu\;e^{-\int_0^Tdt\,\cL}
\end{equation}
and
\begin{equation}
\left\langle\vek{c}^A(T)|e^{-H_{\rm eff}T}|\vek{c}^A(0)\right\rangle=
\int\cD c^A_j  \;e^{-\int_0^Tdt\,\cL_{\rm eff}},
\end{equation}
where $\cL$ and $\cL_{\rm eff}$ are Euclidean Lagrangians.

To the order of interest $\cL_{\rm eff}$ is given by the two- and
four-point diagrams in fig.~\ref{diagrams}.
\begin{figure}[b]
\epsfbox{cpbcII2.eps}
\caption[diagrams]{One loop Feynman diagrams for the two- and four-point
functions.}\label{diagrams}
\end{figure}
We use Feynman perturbation theory to evaluate them.
The familiar Feynman rules apply, but with a few modifications.
The external lines denote wave functions $c_i^A/g_0$, and an $n$-point
function acquires a combinatorial factor $1/n!$.
In general the $c_i^A$ are time or, after Fourier transforming,
frequency dependent; $p_0\vek{c}^A$ corresponds to $\dot{\vek{c}}^A$.
We shall only need terms up to $p_0^2$ in the two-point diagrams, and
the $p_0$-independent piece of the four-point function.
The loop frequency is integrated as usual, but the spatial loop momentum
is summed over the values allowed by the boundary condition.
In particular, the loop momentum is $\vek{p}=2\pi\vek{\nu}/L$ for
$A\in\{1,2,3\}$ and $\vek{p}=2\pi(\vek{\nu}\pm\quart\vek{l})/L$ for
$a\in\{4,5,6,7\}$.
Since $f^{aB8}=0$ the eighth component drops out.

Taking permutation and gauge symmetry into account, the effective
Lagrangian takes the form
\begin{equation}
L\cL_{\rm eff} =
g_0^{-2} (\delta_{ij} - g_0^2 p_{ij}) \half \dot{c}_i^A \dot{c}_j^A
  + V_{\rm eff}(\vek{c}^A),
\end{equation}
where $V_{\rm eff}(\vek{c}^A)$ is given below in eq.~(\ref{hamilton}).
In the Hamiltonian the kinetic term is
$g_0^2(\delta_{ij}+g_0^2p_{ij})\half e_i^Ae_j^A$, or after rescaling
$c_i^A\mapsto g_0^{2/3}c_i^A$ and $e_i^A\mapsto g_0^{-2/3}e_i^A$,
\begin{equation}
\begin{array}{r}
LH_{\rm eff} =
  g_0^{2/3}(\delta_{ij} + g_0^2 p_{ij}) \half e_i^A e_j^A
+ g_0^{2/3}(\delta_{ij} + g_0^2 q_{ij}) \quart F_{ik}^A F_{jk}^A
   \\[1.0em]
+ g_0^{4/3} m_{ij} c_i^A c_j^A
+ g_0^{8/3} \cS_{ijkl} s^{ABDE} c_i^A c_j^B c_k^D c_l^E ,
\end{array}
\end{equation}
where $F_{ij}^A=f^{ABD}c_i^Bc_j^D$, and
\begin{equation}
s^{ABDE}=\frac{2}{3}\left( \delta^{AB} \delta^{DE}+
\delta^{BD} \delta^{EA}+ \delta^{AD} \delta^{BE} \right) .
\end{equation}

In an arbitrary covariant gauge with gluon propagator
\begin{equation}
\frac{1}{p^2}\left(\delta_{\mu\nu} -
(1-\alpha)\frac{p_\mu p_\nu}{p^2}\right)
\end{equation}
explicit evaluation of the Feynman diagrams yields
\begin{equation}
p_{ij} = \frac{1}{8} \left(2 {\sum}'+ {\sum}'' \right)
\left( 2(3-\alpha)\frac{\delta_{ij}}{k^3}
- (d+2-3\alpha)\frac{k_ik_j}{k^5} \right) ,
\end{equation}
\begin{equation}
q_{ij} = - \frac{1}{24} \left(2 {\sum}'+ {\sum}'' \right)
\left( (d+11-12\alpha)\frac{\delta_{ij}}{k^3}
- 6(d+2-3\alpha)\frac{k_ik_j}{k^5} \right) ,
\end{equation}
\begin{equation}
m_{ij} = \frac{(d-1)}{4} \left(2 {\sum}'+ {\sum}'' \right)
\left( \frac{\delta_{ij}}{k} - \frac{k_ik_j}{k^3} \right) ,
\end{equation}
and
\begin{equation}
\cS_{ijkl} = -
\frac{(d-1)}{16} \left({\sum}'+ \frac{1}{8} {\sum}'' \right)
\left( \frac{\delta_{\{ij}\delta_{kl\}}}{k^3}
- 6\frac{k_{\{i}k_j\delta_{kl\}}}{k^5}
+ 5\frac{k_ik_jk_kk_l}{k^7} \right) .
\end{equation}
The notation $\{\cdots\}$ implies complete symmetrization of the
embraced indices.
The summation symbols and their associated loop momentum values are
\begin{equation}
{\sum}'  := \sum_{\scriptstyle\vek{\nu}\neq\veg{0}},\;\;
\vek{k}=2\pi\vek{\nu}
\end{equation}
from the periodic modes in the loops, and
\begin{equation}
{\sum}'' := \sum_{\scriptstyle\vek{\nu}},\;\;
\vek{k}=2\pi(\vek{\nu}-\quart\vek{l})
\end{equation}
from the $\pm i$-periodic modes.
The difference in the sums is an infrared effect;
in particular, the poles as $d\rightarrow3$ are identical, and are
absorbed by the usual SU(3) coupling constant renormalization.

It would have been much too tedious to obtain these results by hand for
arbitrary $\alpha$, so we used the symbolic manipulation language FORM.
We checked the FORM programs in several ways.
For example, it was easy to reproduce the known result for the
infinite-volume two-point function in an arbitrary gauge, since the
Feynman rules for vertices are the same.

Note that $m_{ij}$ and $\cS_{ijkl}$ are gauge invariant, but
$p_{ij}$ and $q_{ij}$ depend on $\alpha$.
However, the wave-function renormalization
\begin{equation}
c_i^a\rightarrow Z_{ij}^{1/2}  c_j^a ,\;\;\;
e_i^a\rightarrow Z_{ij}^{-1/2} e_j^a ,
\end{equation}
where
\begin{equation}
Z_{ij}=\delta_{ij}+g_0^2 z_{ij},\;\;\;
z_{ij} = \third (p_{ij} - q_{ij}) ,
\end{equation}
removes the gauge dependence, and sets the coefficients of the
kinetic term and the tree-level potential term equal.
In addition, after renormalization the $k_ik_j/k^5$ terms cancel:
\begin{equation}
p_{ij}-z_{ij} = q_{ij}+2z_{ij} = \delta_{ij}\alpha_2,
\end{equation}
where
\begin{equation}
\alpha_2 =
\frac{25-d}{72} \left(2 {\sum}'+ {\sum}'' \right) \frac{1}{k^3}=
\frac{11}{3(4\pi)^2}\frac{1}{\varepsilon} + a_2,
\end{equation}
with $d = 3 - 2 \varepsilon$. This is a pleasant surprise.
After this field renormalization and MS-scheme coupling constant
renormalization
\begin{equation}\label{hamilton}
\begin{array}{r}
LH_{\rm eff} =
  g^{2/3}(1 + g^2 a_2)
    \left( \half e_i^A e_i^A + \quart F_{ij}^A F_{ij}^A \right)
\hspace*{1.0cm} \\[1.0em] \hspace*{1.0cm}
  +\, g^{4/3} m_{ij} c_i^A c_j^A
\,+\, g^{8/3} \cS_{ijkl} s^{ABDE} c_i^A c_j^B c_k^D c_l^E ,
\end{array}
\end{equation}
where $g^2=g^2_{\rm MS}(1/L)$.
Numerical values for $a_2$, $m_{ij}$, and $\cS_{ijkl}$ are
in Table~\ref{coefficients}.
\begin{table}
\caption[coefficients]{The coefficients in the effective Hamiltonian.
The tensors $m_{ij}$ and $\cS_{ijkl}$ are completely
symmetric.}\label{coefficients}
\vspace{1.0em}
\begin{center} \begin{tabular}{cr@{.}lr@{.}lr@{.}l}
\hline \hline
          coefficient           & \multicolumn{2}{c}{$\sum'$} &
\multicolumn{2}{c}{$\sum''$} & \multicolumn{2}{c}{total}  \\
\hline
 $3(4\pi)^2a_2/11$   &    $-0$&2727019    &  $0$&5986198 &  $0$&3259179 \\
   $6\pi m_{ii}$     &    $-5$&6745750    & $-0$&2004840 & $-5$&8750590 \\
   $6\pi m_{ij}$     &\multicolumn{2}{c}{}& $-1$&4383214 & $-1$&4383214 \\
$60\pi^2\cS_{iiii}$  &    $-0$&4289976    &  $0$&2502256 & $-0$&1787720 \\
$60\pi^2\cS_{ijjj}$  &\multicolumn{2}{c}{}&  $0$&0446557 &  $0$&0446557 \\
$60\pi^2\cS_{iijj}$  &       0&6311655    & $-0$&0730294 &  $0$&5581361 \\
$60\pi^2\cS_{ijkk}$  &\multicolumn{2}{c}{}& $-0$&0039135 & $-0$&0039135 \\
\hline \hline
\end{tabular} \end{center}
\end{table}
Notice how the quarter-momentum sums lead to terms in $H_{\rm eff}$ that
break the cubic symmetry down to the permutation symmetry, as discussed
in sect.~\ref{degen}.

After renormalization, these formulae straightforwardly reproduce
previous calculations of L\"uscher \cite{Lue83} when the momenta are
color blind.
Moreover, we also checked the final results by repeating the
calculations using L\"uscher's technique of Bloch perturbation theory.

Some qualitative properties of the spectrum follow from
eq.~(\ref{hamilton}) immediately.
To lowest order in $g^{2/3}$, which should pertain to the smallest
of volumes, the effective Hamiltonian is invariant under the full
rotation group O(3).
Indeed, it is identical to the effective Hamiltonian of SU(2) in a
periodic box.
(This is because the quartic modes at any absolute minimum of
$\cV_{\rm eff}$ form an su(2) subalgebra.)
 From ref.~\cite{Lue84} we know, therefore, that the vacuum has quantum
numbers $J^P=0^+$, where $P$ refers to parity.
Furthermore, the $2^+$ glueball lies lower than
the $0^+$ glueball \cite{Lue84}.
For periodic SU(2) all these states (including the vacuum) exhibit an
eight-fold degeneracy, associated with central conjugations.
For $C$-periodic SU(3) they exhibit a four-fold degeneracy, associated
with the rotations connecting the four minima of $\cV_{\rm eff}$.
Of course, in SU(3) one must specify charge conjugation; all states
built out of the quartic modes $\vek{c}^A$ have $C=+$.

At ${\rm O}(g^{4/3})$ the $m_{ij}$-terms in eq.~(\ref{hamilton}) break
the O(3) symmetry down to the permutation group $S_3$.
The $0^{++}$ states are simply labeled as $\IRSaone$ states, but the
$2^{++}$ states split into a $\IRSaone$ and two $\IRSe$ states
with different energies.
Higher order terms respect this pattern.
All states are in addition 4-fold degenerate, but
as discussed in sect.~\ref{degen} this degeneracy is lifted
non-perturbatively by tunneling between different perturbatively
degenerate states.

Altogether one may distinguish four regimes: very small volumes, where
lowest order perturbation theory applies, and the states are labeled by
representations of the full rotation group O(3); somewhat larger
volumes, where higher order perturbative contributions become
non-negligible and the $S_3$ multiplets are resolvable; intermediate
volumes, where tunneling restores the cubic rotation group O(3,$\Z$);
and, finally, large volumes, where the full O(3) rotation group
reappears.
This demonstrates that also with $C$-periodic boundary conditions there is
no simple connection between the small volume perturbative regime and the
large volume non-perturbative regime.

\section{Conclusions}\label{conclude}
A preliminary write-up of these results \cite{Kro90} drew the incorrect
conclusion that the absolute minimum of $\cV_{\rm eff}$ was unique.
If that had been the case, the volume dependence of glueball masses in
a $C$-periodic box may have matched more smoothly onto large volume
results than in a periodic box.
Since we now know that there are {\em four\/} degenerate minima of the
toron effective potential, tunneling transitions are expected.
However, the SU(3) $C$-periodic effective Hamiltonian is similar
(at lowest order identical) to its SU(2) periodic counterpart.
Consequently, it might be simpler to solve than the periodic SU(3)
case \cite{Wei87,Voh88}.
Of course, there may be technical complications owing to the local
minimum.

A complete spectrum calculation goes beyond the scope of this paper,
which concentrates on the derivation of the effective Hamiltonian and
some of its qualitative features. A spectrum calculation
would nevertheless be interesting, especially when confronted with
numerical simulations of the glueball spectrum in a $C$-periodic box.
In simulations one can also probe the physically most interesting large
volume regime.
A comparison of periodic and $C$-periodic systems may reveal how
sensitive the gluon system is to the infrared cutoff, and how large the
volume must be before we see non-perturbative large volume results.

\section*{Acknowledgements}
One of us (A.S.K.) would like to thank the HLRZ, J\"ulich, and the
Institut f\"ur Theoretische Physik, Bern, for hospitality while this
work was being carried out.

\appendix

\section{Torons are Real, Constant and Abelian}\label{proof}

The potential energy is
\begin{equation}\label{YM-potential}
V[\vek{A}]=\frac{1}{4 g_0^2}\int d^dx\,(F^a_{ij})^2,
\end{equation}
where $0\leq x_i<L$,
\begin{equation}
F_{ij}=\nabla_i A_j - \nabla_j A_i + [A_i, A_j]=F^a_{ij}T^a
\end{equation}
is the field strength, and $g_0$ is the bare coupling.
Obviously, if $\vek{A}$ is constant ($\nabla_iA_j=0$) and abelian
($[A_i,A_j]=0$), $F_{ij}$ and $V$ vanish.
Ref.~\cite{Kro91} showed that any $C$-periodic boundary condition is
gauge-equivalent to eq.~(\ref{component-bc}).
With those boundary conditions, however, a constant $\vek{A}$ must
also be real.
On the other hand, suppose the field strength (and therefore the
potential energy) vanishes for some $\vek{A}$.
We want to show that $\vek{A}$ is gauge equivalent to a real, constant,
abelian configuration.

Let $a$ be the gauge transformation that brings $\vek{A}$ into
the complete axial gauge:
\begin{equation}\label{axial-gauge}
\begin{array}{ll}
^a\!A_1(\vek{x}) = 0 & \mbox{everywhere,}      \\
^a\!A_2(\vek{x}) = 0 & \mbox{for $x_1=0$,}     \\
^a\!A_3(\vek{x}) = 0 & \mbox{for $x_1=x_2=0$,} \ldots.
\end{array}
\end{equation}
This gauge transformation induces non-trivial transition functions
$^{a}\Omega_i$.
The field \linebreak
strength vanishes everywhere if and only if parallel transport
around any contractable path is trivial.
Starting with infinitesimal paths near $\vek{x}=\veg{0}$, it is easy to
see that this is so only if $^a\!\vek{A}(\vek{x})=0$ and only if
$^{a}\Omega_i(\vek{x})=V_i$, where $V_i$ are constant matrices obeying
\begin{equation}\label{VV}
V_i V_j^* = V_j V_i^*.
\end{equation}
The condition on the $^{a}\Omega_i$ comes from considering paths that
cross two boundaries but are still contractable.

There is a gauge transformation $b$, constructed in Appendix B, such that
\begin{equation}\label{bV}
^{ba}\Omega_i=\,^bV_i=\exp(T^2C_i),
\end{equation}
but still preserving the complete axial gauge.
Now let $c=\exp(-T^2\vek{C}\cdot\vek{x}/L)$ and $g=cba$.
Then $^g\vek{A}=\vek{C}T^2/L$ is real, constant and abelian as claimed.
Moreover, the transition functions have been gauged away again,
$^g\Omega_i=1$.

Carrying out the gauge transformation $h$ in eq.~(\ref{C-shift}) shifts
the value of $\vek{C}$, viz.\ $^h\vek{C}=\vek{C}-\pi\vek{l}$.
Hence, under the boundary condition of eq.~(\ref{cpbcO}) the torons are
real, constant and abelian as well.

\section{Construction of $b$}
The axial gauge condition does not fix gauge transformations that are
constant for $0\leq x_i<L$, but complex conjugated each time
$x_i\geq L$.
Under such a transformation $v:\;V_i\mapsto vV_iv^T$.
Let $W_i=V_iV_i^*$, and note that repeated use of eq.~(\ref{VV}) implies
that all $W_i$ can be simultaneously diagonalized.
Denote the unitary matrix that carries out the diagonalization by $u$,
and write
\begin{equation}
\tilde{V}_i=uV_iu^T,\;\;\;
\tilde{V}_i \tilde{V}_i^*=\Lambda_i=
\diag(\lambda_i^{(1)}, \lambda_i^{(2)}, \lambda_i^{(3)}).
\end{equation}
Then
\begin{equation}\label{VLambda}
\tilde{V}_i=\Lambda_i\tilde{V}_i^{\rm T}=
(\tilde{V}_i\Lambda_i)^{\rm T},
\end{equation}
which implies, for each $i$, that $\tilde{V}_i$ falls into one of two
cases:
Case (0): $\Lambda_i$ is the unit matrix.
Then $\tilde{V}_i=\tilde{V}_i^{\rm T}$ and hence $V_i=V_i^{\rm T}$.
Case ($\alpha,\beta$): $\lambda_i^{(\alpha)}=e^{i2C_i}$, $C_i\neq0$,
for some $\alpha$.
Then eq.~(\ref{VLambda}) says that the other two entries of $\Lambda_i$
are $\lambda_i^{(\beta)}=e^{-i2C_i}$ and $\lambda_i^{(\gamma)}=1$, and
it says that all elements of $\tilde{V}_i$ vanish except
$\tilde{V}_i^{(\alpha\beta)}=ie^{-i(C_i-\Phi)}$,
$\tilde{V}_i^{(\beta\alpha)}=ie^{i(C_i+\Phi)}$,
$\tilde{V}_i^{(\gamma\gamma)}=e^{-i2C_i}$.

Either there is a direction $I$ such that $\tilde{V}_I$ falls under
case (0), or all $\tilde{V}_i$ fall under a case ($\alpha,\beta$).
Suppose the former, i.e.\ $V_I^{}=V_I^{\rm T}$.
Since $V_I\in{\rm SU}(N)$ there is a matrix $w$ such that
$w^2=V_I^{-1}$, and $w$ is symmetric, since $V_I$ is.
Now define the $C$-periodic gauge transformation $b_1$ to be
$w$ for $0\leq x_i<L$, $w^*$ if one $x_i>L$, etc.
Then $^{b_1}V_I=wV_Iw^{\rm T}=1$ and eq.~(\ref{VV}) implies that all
$^wV_j$ are real.
But if all $V_i$ were real, eq.~(\ref{VV}) would imply that they
commute.
Define $C_i$ by $^{b_1}V_i=\exp(\omega^A T^A C_i)$, where
$A\in\{2,5,7\}$.
This is possible because the $^{b_1}V_i$ are real and commute.
Here
\begin{equation}
\begin{array}{l}
\omega^2 = \cos\theta, \\[0.5em]
\omega^5 = \sin\theta\,\cos\phi, \\[0.5em]
\omega^7 = \sin\theta\,\sin\phi
\end{array}
\end{equation}
form a unit vector.
The real, constant gauge transformation
\begin{equation}
b_2=\left(
\begin{array}{ccc}
\cos\phi             & -\sin\phi            & 0 \\
\cos\theta\,\sin\phi & \cos\theta\,\cos\phi & -\sin\theta \\
\sin\theta\,\sin\phi & \sin\theta\,\cos\phi &  \cos\theta
\end{array}
\right)
\end{equation}
rotates $\omega^A T^A$ to $T^2$.
Hence, setting $b=b_2b_1$, $^bV_i=\exp(T^2C_i)$ as in eq.~(\ref{bV}).

Now suppose all $\tilde{V}_i$ fall under a case ($\alpha,\beta$).
Eq.~(\ref{VV}) requires that $\Phi$, $\alpha$, and
$\beta$ be the same for all $i$.
Without loss one may assume $(\alpha,\beta)=(1,2)$.
Then define
\begin{equation}
v=\frac{1}{\sqrt{2}}\left(
\begin{array}{ccc}
  e^{-i\Phi/2} & i e^{-i\Phi/2} & 0 \\
i e^{-i\Phi/2} &   e^{-i\Phi/2} & 0 \\
      0        &       0        & \sqrt{2} e^{i\Phi}
\end{array}
\right) .
\end{equation}
The gauge transformation that brings $V_i$ into the desired form is
$b=vu$ for $0\leq x_i<L$, complex conjugated each time $x_j>L$.


\begin{thebibliography}{99}
\bibitem{Kro91} 
A.S. Kronfeld and U.-J. Wiese, Nucl.\ Phys.\ B357 (1991) 521
\bibitem{Pol91} %
L. Polley and U.-J. Wiese, Nucl.\ Phys.\ B356 (1991) 629
\bibitem{Wie91} 
U.-J. Wiese, Nucl.\ Phys.\ B375 (1992) 45
\bibitem{Lue83} 
M. L\"uscher, Nucl.\ Phys.\ B219 (1983) 233
\bibitem{tHo79} %
G. 't~Hooft, Nucl.\ Phys.\ B153 (1979) 141
\bibitem{Kol86} %
J. Koller and P. van Baal, Nucl.\ Phys.\ B273 (1986) 387;
Ann.\ Phys.\ (New York) 174 (1987) 299;
Nucl.\ Phys.\ B302 (1988) 1
\bibitem{vBa88} 
P. van Baal, Nucl.\ Phys.\ B307 (1988) 274; B312 (1889) 752
\bibitem{Lue84} 
M. L\"uscher and G. M\"unster, Nucl. Phys. B232 (1984) 445
\bibitem{Wei87} 
P. Weisz and V. Ziemann, Nucl. Phys. B284 (1987) 157
\bibitem{Voh88} 
C. Vohwinkel, Phys. Lett. B213 (1988) 54; Phys. Rev. Lett. 63 (1989) 2544
\bibitem{Kro90} 
A.S. Kronfeld, Nucl.\ Phys.\ B (Proc.\ Suppl.) 20 (1991) 20
\end{thebibliography}
\end{document}